\documentclass[english,nofootinbib,showpacs]{revtex4}
\usepackage[T1]{fontenc}
\usepackage[latin9]{inputenc}
\usepackage{float}
\usepackage{amsmath}
\usepackage{amssymb}
\usepackage{graphicx}
\usepackage{esint}

\makeatletter

\@ifundefined{textcolor}{}
{%
 \definecolor{BLACK}{gray}{0}
 \definecolor{WHITE}{gray}{1}
 \definecolor{RED}{rgb}{1,0,0}
 \definecolor{GREEN}{rgb}{0,1,0}
 \definecolor{BLUE}{rgb}{0,0,1}
 \definecolor{CYAN}{cmyk}{1,0,0,0}
 \definecolor{MAGENTA}{cmyk}{0,1,0,0}
 \definecolor{YELLOW}{cmyk}{0,0,1,0}
}

\makeatother

\usepackage{babel}
\begin{document}

\title{A hyperelliptic solution class for the hyperbolic Ernst equation}

\author{Sebastian Moeckel}

\affiliation{Theoretisch Physikalisches Institut, Friedrich-Schiller-Universität
Jena, Max-Wien-Platz 1, 07743 Jena, Germany}

\email{sebastian.moeckel@uni-jena.de}

\selectlanguage{english}%
\begin{abstract}
\noindent A new hyperelliptic solution class for the hyperbolic Ernst
equation is obtained by transforming the regarding solution of the
elliptic Ernst equation. Furthermore, a nontrivial way for obtaining
general polarized colliding wave solutions from this hyperelliptic
family of solutions is presented. The explicit form of the solutions
for a Riemann surface of genus $n=1$ is given. In addition, an explicit
example in terms of a Khan-Penrose seed is provided, emphasizing the
importance of the presented procedure for generating general polarized
colliding plane-wave space times from space-times with a collinear
polarization of the colliding waves.
\end{abstract}


\maketitle

\section{Introduction}

The search for new, inherently non-linear solutions of the hyperbolic
Ernst equation, gains importance particularly in the context of colliding
plane wave space-times. Here, the similarity between the elliptic
and hyperbolic Ernst equations in certain coordinate frames can be
employed for easily transforming already known solutions to new ones.
Of course, the knowledge of the underlying symmetry algebra of the
corresponding equations is of uttermost importance for the transformation
procedure. In this paper the hyperelliptic solution class, that has
been found by Meinel and Neugebauer (\cite{meinel1996solutions})
for the elliptic Ernst equation, will be transformed into a solution
class of the hyperbolic Ernst equation. Furthermore, an outlook on
utilizing this solution class for generating colliding plane wave
space-times with non-collinear polarization of the colliding waves
is provided.

\section{The hyperbolic and elliptic Ernst equations}

The elliptic Ernst equation has first been considered by Ernst (\cite{ernst1968new})
in the context of stationary axisymmetric space-times and can be written
in polar coordinates $\left(\rho,\zeta\right)$ as follows:
\begin{eqnarray}
\left(\Re\hat{Z}\right)\nabla^{2}\hat{Z} & = & \left(\nabla\hat{Z}\right)^{2},\label{eq:Ernst_ell}
\end{eqnarray}
where $\hat{Z}:\;\mathbb{R}\times\mathbb{R}\rightarrow\mathbb{C}$,
$\nabla=\left(\frac{\partial}{\partial\rho},\frac{\partial}{\partial\zeta}\right)$
and $\nabla^{2}=\frac{\partial^{2}}{\partial\rho^{2}}+\frac{1}{\rho}\frac{\partial}{\partial\rho}+\frac{\partial^{2}}{\partial\zeta^{2}}$.
Transforming (\ref{eq:Ernst_ell}) to complex coordinates $\left(z,\bar{z}\right)$,
given by
\begin{align}
z=\rho+i\zeta, & \quad & \bar{z}=\rho-i\zeta,\label{eq:z_zb}
\end{align}
yields the following form of the elliptic Ernst equation
\begin{eqnarray}
\left(\hat{Z}+\bar{\hat{Z}}\right)\left[2\hat{Z}_{z\bar{z}}+\frac{\hat{Z}_{z}+\hat{Z}_{\bar{z}}}{z+\bar{z}}\right] & = & 4\hat{Z}_{z}\hat{Z}_{\bar{z}}\label{eq:z_zq_ernst}
\end{eqnarray}
In full analogy, the hyperbolic Ernst equation can be written in the
following form
\begin{eqnarray}
\left(Z+\bar{Z}\right)\left[2Z_{fg}+\frac{Z_{f}+Z_{g}}{f+g}\right] & = & 4Z_{f}Z_{g}\label{eq:f_g_hyp_ernst}
\end{eqnarray}
where $Z:\;\mathbb{R}\times\mathbb{R}\rightarrow\mathbb{C}$ is a
complex function of the two independent real coordinates $\left(f,g\right)$.\\
According to \cite{mockel2013solutions}, the Ernst equation (\ref{eq:f_g_hyp_ernst})
admits the following point symmetries
\begin{eqnarray}
Z\left(f,g\right) & \overset{}{\longrightarrow} & Z\left(f+\beta,g-\beta\right),\label{eq:K_L_trans}\\
Z\left(f,g\right) & \overset{}{\longrightarrow} & Z\left(e^{\alpha}f,e^{\alpha}g\right),\label{eq:K_L_shift_arg}\\
Z\left(f,g\right) & \overset{}{\longrightarrow} & Z\left(f,g\right)+\gamma,\label{eq:L_shift}\\
Z\left(f,g\right) & \overset{}{\longrightarrow} & e^{\delta}Z\left(f,g\right),\label{eq:K_L_scale}
\end{eqnarray}
where $\alpha,\beta,\gamma,\delta\in\mathbb{C}$. Of course, the elliptic
equation (\ref{eq:z_zq_ernst}) exhibits the same symmetries.

\section{The hyperelliptic solution class}

\noindent The hyperelliptic solution class has been first considered
by Meinel and Neugebauer (\cite{meinel1996solutions}). Accordingly,
it has been shown that a solution class in terms of hyperelliptic
integrals can be obtained by considering solutions of the stationary
axisymmetric vacuum field equations associated with Jacobi's inversion
problem
\begin{eqnarray}
\hat{Z} & = & \exp\left(\sum_{m=1}^{n}\intop_{\kappa_{m}}^{\kappa^{\left(m\right)}}\frac{\kappa^{n}d\kappa}{\hat{W}}-u_{n}\right),\label{eq:meinel_hyperelliptic1}
\end{eqnarray}
where 
\begin{eqnarray}
\hat{W}\left(\kappa;z,\bar{z}\right){}^{2} & = & \left(\kappa+iz\right)\left(\kappa-i\bar{z}\right)\prod_{j=1}^{n}\left(\kappa-\kappa_{j}\right)\left(\kappa-\bar{\kappa}_{j}\right),\label{eq:riemann_W}
\end{eqnarray}
and $\kappa_{j}$ $\left(j=1,\ldots,n\right)$ are arbitrary complex
constants. Furthermore, the upper integration limits in (\ref{eq:meinel_hyperelliptic1})
are functions of $z$ and $\bar{z}$, which have to be calculated
by solving the following inversion problem
\begin{alignat}{3}
\sum_{m=1}^{n}\intop_{\kappa_{m}}^{\kappa^{\left(m\right)}}\frac{\kappa^{j}d\kappa}{\hat{W}} & = & u_{j}, & \quad & j=0,1,2,\ldots,n-1,\label{eq:inversion_prob}
\end{alignat}
where the $u_{j}$ are real functions of $z$ and $\bar{z}$, which
solve the Euler-Poisson-Darboux (EPD) equation
\begin{eqnarray}
2\left(z+\bar{z}\right)\partial_{z\bar{z}}^{2}u_{j}+\partial_{z}u_{j}+\partial_{\bar{z}}u_{j} & = & 0.\label{eq:EPD_uj}
\end{eqnarray}
In addition, the $u_{j}$ are required to satisfy the following recursive
relations
\begin{alignat}{3}
i\partial_{z}u_{j} & = & \frac{1}{2}u_{j-1}+z\partial_{z}u_{j-1}, & \quad & j=1,2,\ldots,n,\label{eq:rec_uj}\\
-i\partial_{\bar{z}}u_{j} & = & \frac{1}{2}u_{j-1}+\bar{z}\partial_{\bar{z}}u_{j-1}, & \quad & j=1,2,\ldots,n.\label{eq:rec_uj2}
\end{alignat}
Note, that the EPD equation for $u_{j-1}$ occurs as the integrability
condition for the system (\ref{eq:rec_uj})-(\ref{eq:rec_uj2}).

\subsection*{\noindent Adapting the hyperelliptic class to the hyperboliv Ernst
equation}

\noindent Some requirements have to be met for transforming the solution
(\ref{eq:meinel_hyperelliptic1}) of the elliptic Ernst equation to
a new solution of the hyperbolic Ernst equation. Note, that the proof
in \cite{meinel1996solutions} of (\ref{eq:meinel_hyperelliptic1})
constituting a solution of the elliptic Ernst equation relies essentially
on three pillars: The functions $u_{j}$ defined via (\ref{eq:EPD_uj})-(\ref{eq:rec_uj2})
need to be real-valued, and the relations 
\begin{eqnarray}
\sum_{m=1}^{n}\frac{\kappa^{\left(m\right)}+iz}{\hat{W}^{\left(m\right)}}\left(\kappa^{\left(m\right)}\right)^{j-1}\partial_{z}\kappa^{\left(m\right)} & = & 0,\quad j=1,2,\ldots,n-1,\label{eq:sum_z=00003D0}\\
\sum_{m=1}^{n}\frac{\kappa^{\left(m\right)}-i\bar{z}}{\hat{W}^{\left(m\right)}}\left(\kappa^{\left(m\right)}\right)^{j-1}\partial_{\bar{z}}\kappa^{\left(m\right)} & = & 0,\quad j=1,2,\ldots,n-1,\label{eq:sum_zs=00003D0}
\end{eqnarray}
where $\hat{W}^{\left(m\right)}=\hat{W}\left(\kappa^{\left(m\right)}\right)$,
and 
\begin{eqnarray}
\partial_{z}\ln\hat{Z} & = & \sum_{m=1}^{n}\frac{\kappa^{\left(m\right)}+iz}{\hat{W}^{\left(m\right)}}\left(\kappa^{\left(m\right)}\right)^{n-1}\partial_{z}\kappa^{\left(m\right)},\label{eq:Dz_lnZ}\\
\partial_{\bar{z}}\ln\hat{Z} & = & \sum_{m=1}^{n}\frac{\kappa^{\left(m\right)}-i\bar{z}}{\hat{W}^{\left(m\right)}}\left(\kappa^{\left(m\right)}\right)^{n-1}\partial_{\bar{z}}\kappa^{\left(m\right)},\label{eq:Dzs_lnZ}
\end{eqnarray}
are required to hold. Hence, we have to consider the class of transformations
relating $\left(z,\bar{z}\right)$ to $\left(f,g\right)$, which leave
the Ernst equation (\ref{eq:z_zq_ernst}) and these three conditions
invariant. It turns out that the following transformation meets all
the requirements
\begin{eqnarray}
z & \rightarrow & i\left(\alpha f+\beta\right),\label{eq:z_f_g}\\
\bar{z} & \rightarrow & i\left(\alpha g-\beta\right),\label{eq:zs_f_g}\\
Z\left(f,g\right) & = & \hat{Z}\left(i\left(\alpha f+\beta\right),i\left(\alpha g-\beta\right)\right),\label{eq:Z_trafo}\\
W\left(\kappa;f,g\right) & = & \hat{W}\left(\kappa;i\left(\alpha f+\beta\right),i\left(\alpha g-\beta\right)\right),\label{eq:W_trafo}
\end{eqnarray}
where $\alpha\neq0$, $\beta\in\mathbb{R}$.\\
Accordingly, a new solution class for the hyperbolic Ernst equation
has been generated 
\begin{eqnarray}
Z & = & \exp\left(\sum_{m=1}^{n}\intop_{\kappa_{m}}^{\kappa^{\left(m\right)}}\frac{\kappa^{n}d\kappa}{W}-u_{n}\right),\label{eq:meinel_sol_hyper}
\end{eqnarray}
where 
\begin{eqnarray}
W\left(\kappa;f,g\right){}^{2} & = & \left(\kappa-\alpha f-\beta\right)\left(\kappa+\alpha g-\beta\right)\prod_{j=1}^{n}\left(\kappa-\kappa_{j}\right)\left(\kappa-\bar{\kappa}_{j}\right),\label{eq:riemann_W-2}
\end{eqnarray}
and $\kappa_{j}$ $\left(j=1,\ldots,n\right)$ are arbitrary complex
constants. In full analogy to the elliptic case, the upper integration
limits in (\ref{eq:meinel_hyperelliptic1}) are functions of $f$
and $g$, which have to be calculated by solving the following inversion
problem
\begin{alignat}{3}
\sum_{m=1}^{n}\intop_{\kappa_{m}}^{\kappa^{\left(m\right)}}\frac{\kappa^{j}d\kappa}{W} & = & u_{j}, & \quad & j=0,1,2,\ldots,n-1\label{eq:inversion_prob-2}
\end{alignat}
where the $u_{j}$ are real functions of $f$ and $g$, which solve
the EPD equation%
\footnote{\noindent The EPD equation is invariant under the transformation (\ref{eq:z_f_g})-(\ref{eq:zs_f_g}),
cf. \cite{miller1973symmetries}.%
}
\begin{eqnarray}
2\left(f+g\right)\partial_{fg}^{2}u_{j}+\partial_{f}u_{j}+\partial_{g}u_{j} & = & 0.\label{eq:EPD_uj-2}
\end{eqnarray}
In addition, the $u_{j}$ are required to satisfy the following recursive
relations
\begin{alignat}{3}
\partial_{f}u_{j} & = & \frac{\alpha}{2}u_{j-1}+\left(\alpha f+\beta\right)\partial_{f}u_{j-1}, & \quad & j=1,2,\ldots,n,\label{eq:rec_uj-1}\\
\partial_{g}u_{j} & = & -\frac{\alpha}{2}u_{j-1}-\left(\alpha g-\beta\right)\partial_{g}u_{j-1}, & \quad & j=1,2,\ldots,n.\label{eq:rec_uj2-1}
\end{alignat}
Therefore, the $u_{j}$ are certainly real-valued for $\alpha$, $\beta\in\mathbb{R}$,
if $u_{0}$ is real-valued.\\
Consequently, the proof of the invariance of relations (\ref{eq:sum_z=00003D0})-(\ref{eq:sum_zs=00003D0})
proceeds as follows. Differentiating (\ref{eq:inversion_prob-2})
with respect to $f$ and $g$, yields the important intermediate result:
\begin{alignat}{3}
\sum_{m=1}^{n}\left\{ \frac{\left(\kappa^{\left(m\right)}\right)^{j}}{W^{\left(m\right)}}\partial_{f}\kappa^{\left(m\right)}+\frac{\alpha}{2}\intop_{\kappa_{m}}^{\kappa^{\left(m\right)}}\frac{\kappa^{j}d\kappa}{\left(\kappa-\alpha f-\beta\right)W}\right\}  & = & \partial_{f}u_{j}, & \quad & j=0,1,2,\ldots,n-1,\label{eq:Dz_uj-1}\\
\sum_{m=1}^{n}\left\{ \frac{\left(\kappa^{\left(m\right)}\right)^{j}}{W^{\left(m\right)}}\partial_{g}\kappa^{\left(m\right)}-\frac{\alpha}{2}\intop_{\kappa_{m}}^{\kappa^{\left(m\right)}}\frac{\kappa^{j}d\kappa}{\left(\kappa+\alpha g-\beta\right)W}\right\}  & = & \partial_{g}u_{j}, & \quad & j=0,1,2,\ldots,n-1,\label{eq:Dzs_uj-1}
\end{alignat}
where $W^{\left(m\right)}=W\left(\kappa^{\left(m\right)}\right)$
and the following identities have been used
\begin{eqnarray}
\partial_{f}\frac{1}{W} & = & \frac{\alpha}{2\left(\kappa-\alpha f-\beta\right)W},\label{eq:Df_1/w}\\
\partial_{g}\frac{1}{W} & = & -\frac{\alpha}{2\left(\kappa+\alpha g-\beta\right)W}.\label{eq:Dg_1/W}
\end{eqnarray}
Hence, the transformed relations (\ref{eq:sum_z=00003D0})-(\ref{eq:sum_zs=00003D0})
follow immediately
\begin{eqnarray*}
\partial_{f}u_{j} & = & \sum_{m=1}^{n}\left\{ \frac{\left(\kappa^{\left(m\right)}\right)^{j}}{W^{\left(m\right)}}\partial_{f}\kappa^{\left(m\right)}+\frac{\alpha}{2}\intop_{\kappa_{m}}^{\kappa^{\left(m\right)}}\frac{\kappa^{j}d\kappa}{\left(\kappa-\alpha f-\beta\right)W}\right\} \\
 & \overset{(\ref{eq:rec_uj-1})}{=} & \left(\alpha f+\beta\right)\partial_{f}u_{j-1}+\frac{\alpha}{2}u_{j-1}\\
 & = & \left(\alpha f+\beta\right)\sum_{m=1}^{n}\left\{ \frac{\left(\kappa^{\left(m\right)}\right)^{j-1}}{W^{\left(m\right)}}\partial_{f}\kappa^{\left(m\right)}+\frac{\alpha}{2}\intop_{\kappa_{m}}^{\kappa^{\left(m\right)}}\frac{\kappa^{j-1}d\kappa}{\left(\kappa-\alpha f-\beta\right)W}\right\} \\
 &  & +\frac{\alpha}{2}\sum_{m=1}^{n}\intop_{\kappa_{m}}^{\kappa^{\left(m\right)}}\frac{\kappa^{j-1}d\kappa}{W}
\end{eqnarray*}
And by equating both expressions one obtains
\begin{eqnarray*}
 &  & \sum_{m=1}^{n}\frac{\kappa^{\left(m\right)}-\alpha f-\beta}{W^{\left(m\right)}}\left(\kappa^{\left(m\right)}\right)^{j-1}\partial_{f}\kappa^{\left(m\right)}\\
 &  & =\frac{\alpha}{2}\sum_{m=1}^{n}\intop_{\kappa_{m}}^{\kappa^{\left(m\right)}}\underbrace{\left\{ \frac{\kappa^{j-1}}{W}-\frac{\kappa^{j}}{\left(\kappa-\alpha f-\beta\right)W}+\frac{\left(\alpha f+\beta\right)\kappa^{j-1}}{\left(\kappa-\alpha f-\beta\right)W}\right\} }_{=\frac{1}{\left(\kappa-\alpha f-\beta\right)W}\left(\kappa^{j}-\left(\alpha f+\beta\right)\kappa^{j-1}-\kappa^{j}+\left(\alpha f+\beta\right)\kappa^{j-1}\right)=0}d\kappa
\end{eqnarray*}
which yields
\begin{eqnarray}
\sum_{m=1}^{n}\frac{\kappa^{\left(m\right)}-\alpha f-\beta}{W^{\left(m\right)}}\left(\kappa^{\left(m\right)}\right)^{j-1}\partial_{f}\kappa^{\left(m\right)} & = & 0,\quad j=1,2,\ldots,n-1,\label{eq:sum_z=00003D0-1}
\end{eqnarray}
and similarly
\begin{eqnarray}
\sum_{m=1}^{n}\frac{\kappa^{\left(m\right)}+\alpha g-\beta}{W^{\left(m\right)}}\left(\kappa^{\left(m\right)}\right)^{j-1}\partial_{g}\kappa^{\left(m\right)} & = & 0,\quad j=1,2,\ldots,n-1.\label{eq:sum_zb=00003D0-1}
\end{eqnarray}
Furthermore, considering the first derivative yields
\begin{eqnarray}
\partial_{f}\ln Z & = & \sum_{m=1}^{n}\left\{ \frac{\left(\kappa^{\left(m\right)}\right)^{n}}{W^{\left(m\right)}}\partial_{f}\kappa^{\left(m\right)}+\frac{\alpha}{2}\intop_{\kappa_{m}}^{\kappa^{\left(m\right)}}\frac{\kappa^{n}d\kappa}{\left(\kappa-\alpha f-\beta\right)W}\right\} -\partial_{f}u_{n}\nonumber \\
 & \overset{\left(\ref{eq:rec_uj-1}\right)}{=} & \sum_{m=1}^{n}\left\{ \frac{\left(\kappa^{\left(m\right)}\right)^{n}}{W^{\left(m\right)}}\partial_{f}\kappa^{\left(m\right)}+\frac{\alpha}{2}\intop_{\kappa_{m}}^{\kappa^{\left(m\right)}}\frac{\kappa^{n}d\kappa}{\left(\kappa-\alpha f-\beta\right)W}\right\} -\frac{\alpha}{2}u_{n-1}-\left(\alpha f+\beta\right)\partial_{f}u_{n-1}\nonumber \\
 & \overset{\left(\ref{eq:Dz_uj-1}\right)}{=} & \sum_{m=1}^{n}\left\{ \frac{\left(\kappa^{\left(m\right)}\right)^{n}}{W^{\left(m\right)}}\partial_{f}\kappa^{\left(m\right)}+\frac{\alpha}{2}\intop_{\kappa_{m}}^{\kappa^{\left(m\right)}}\frac{\kappa^{n}d\kappa}{\left(\kappa-\alpha f-\beta\right)W}\right\} -\frac{\alpha}{2}\sum_{m=1}^{n}\intop_{\kappa_{m}}^{\kappa^{\left(m\right)}}\frac{\kappa^{n-1}d\kappa}{W}\nonumber \\
 &  & -\left(\alpha f+\beta\right)\sum_{m=1}^{n}\left\{ \frac{\left(\kappa^{\left(m\right)}\right)^{n-1}}{W^{\left(m\right)}}\partial_{f}\kappa^{\left(m\right)}+\frac{\alpha}{2}\intop_{\kappa_{m}}^{\kappa^{\left(m\right)}}\frac{\kappa^{n-1}d\kappa}{\left(\kappa-\alpha f-\beta\right)W}\right\} \nonumber \\
 & = & \sum_{m=1}^{n}\frac{\kappa^{\left(m\right)}-\alpha f-\beta}{W^{\left(m\right)}}\left(\kappa^{\left(m\right)}\right)^{n-1}\partial_{f}\kappa^{\left(m\right)}\nonumber \\
 &  & -\frac{\alpha}{2}\sum_{m=1}^{n}\intop_{\kappa_{m}}^{\kappa^{\left(m\right)}}\underbrace{\left[\frac{\kappa^{n-1}}{W}-\frac{\kappa^{n}}{\left(\kappa-\alpha f-\beta\right)W}+\frac{\left(\alpha f+\beta\right)\kappa^{n-1}}{\left(\kappa-\alpha f-\beta\right)W}\right]}_{=\frac{1}{\left(\kappa-\alpha f-\beta\right)W}\left(\kappa^{n}-\left(\alpha f+\beta\right)\kappa^{n-1}-\kappa^{n}+\left(\alpha f+\beta\right)\kappa^{n-1}\right)=0}d\kappa\nonumber \\
 & = & \sum_{m=1}^{n}\frac{\kappa^{\left(m\right)}-\alpha f-\beta}{W^{\left(m\right)}}\left(\kappa^{\left(m\right)}\right)^{n-1}\partial_{f}\kappa^{\left(m\right)},\label{eq:Dz_lnZ-1}
\end{eqnarray}
and analogous
\begin{eqnarray}
\partial_{g}\ln Z & = & \quad\sum_{m=1}^{n}\frac{\kappa^{\left(m\right)}+\alpha g-\beta}{W^{\left(m\right)}}\left(\kappa^{\left(m\right)}\right)^{n-1}\partial_{g}\kappa^{\left(m\right)}.\label{eq:Dzs_lnZ-1}
\end{eqnarray}
Thus, the relations (\ref{eq:sum_z=00003D0})-(\ref{eq:sum_zs=00003D0})
and (\ref{eq:Dz_lnZ})-(\ref{eq:Dzs_lnZ}) still hold, where $iz$
and $-i\bar{z}$ have been replaced by $-\alpha f-\beta$ and $\alpha g-\beta$,
respectively. Now, by following exactly the same steps of the proof
provided in \cite{meinel1996solutions}, it can be shown that (\ref{eq:meinel_sol_hyper})
constitutes a solution of the hyperbolic Ernst equation (\ref{eq:f_g_hyp_ernst}).

\section{Colliding plane wave solutions from the hyperelliptic solution class}

Colliding plane wave space-times emerge as special solution class
of the hyperbolic Ernst equation, satisfying a special type of boundary
conditions on two null surfaces of the underlying manifold. It is
possible to write these wave-conditions in the form of two simple
limit-processes (cf. \cite{griffiths1991colliding}), namely
\begin{alignat}{2}
\frac{1}{4}\leq & \lim_{\left(f,g\right)\rightarrow\left(1/2,1/2\right)}\left[\left(\frac{1}{2}-f\right)\frac{Z_{f}\bar{Z}_{f}}{\left(Z+\bar{Z}\right)^{2}}\right]= & \frac{k_{1}}{2}<\frac{1}{2},\label{eq:boundary2}\\
\frac{1}{4}\leq & \lim_{\left(f,g\right)\rightarrow\left(1/2,1/2\right)}\left[\left(\frac{1}{2}-g\right)\frac{Z_{g}\bar{Z}_{g}}{\left(Z+\bar{Z}\right)^{2}}\right]= & \frac{k_{2}}{2}<\frac{1}{2},\label{eq:boundary1}
\end{alignat}

\noindent For examining the wave conditions in dependence of the functions
$u_{j}$, the first derivatives of $Z$ in terms of the derivatives
of the $u_{j}$ are needed. After some rather lengthy manipulations,
that can be found in the appendix, $\partial_{f}\ln Z$ and $\partial_{g}\ln Z$
read 
\begin{eqnarray}
\partial_{f}\ln Z & = & A\left(f,g\right)\partial_{f}u_{0}\left(f,g\right)+B\left(f,g\right),\label{eq:ansatz_DlnZ_f}\\
\partial_{g}\ln Z & = & C\left(f,g\right)\partial_{g}u_{0}\left(f,g\right)+D\left(f,g\right),\label{eq:ansatz_DlnZ_g}
\end{eqnarray}
where
\begin{eqnarray}
A\left(f,g\right) & = & \sum_{k,i=1}^{n}\left(\kappa^{\left(k\right)}-\alpha f-\beta\right)\left(\kappa^{\left(k\right)}\right)^{n-1}F_{i,k}\left(\kappa^{\left(1\right)},\ldots,\kappa^{\left(n\right)}\right)\left(\alpha f+\beta\right)^{i-1},\label{eq:A(f,g)}\\
B\left(f,g\right) & = & \frac{\alpha}{2}\sum_{k,i=1}^{n}\left(\kappa^{\left(k\right)}-\alpha f-\beta\right)\left(\kappa^{\left(k\right)}\right)^{n-1}F_{i,k}\left(\kappa^{\left(1\right)},\ldots,\kappa^{\left(n\right)}\right)\nonumber \\
 &  & \times\left\{ \sum_{l=1}^{i-1}\left(\alpha f+\beta\right)^{l-1}u_{i-l-1}-\sum_{m=1}^{n}\intop_{\kappa_{m}}^{\kappa^{\left(m\right)}}\frac{\kappa^{i-1}d\kappa}{\left(\kappa-\alpha f-\beta\right)W}\right\} ,\label{eq:B(f,g)}\\
C\left(f,g\right) & = & \sum_{k,i=1}^{n}\left(\kappa^{\left(k\right)}+\alpha g-\beta\right)\left(\kappa^{\left(k\right)}\right)^{n-1}F_{i,k}\left(\kappa^{\left(1\right)},\ldots,\kappa^{\left(n\right)}\right)\left(\beta-\alpha g\right)^{i-1},\label{eq:C(f,g)}\\
D\left(f,g\right) & = & -\frac{\alpha}{2}\sum_{k,i=1}^{n}\left(\kappa^{\left(k\right)}+\alpha g-\beta\right)\left(\kappa^{\left(k\right)}\right)^{n-1}F_{i,k}\left(\kappa^{\left(1\right)},\ldots,\kappa^{\left(n\right)}\right)\nonumber \\
 &  & \times\left\{ \sum_{l=1}^{i-1}\left(\beta-\alpha g\right)^{l-1}u_{i-l-1}-\sum_{m=1}^{n}\intop_{\kappa_{m}}^{\kappa^{\left(m\right)}}\frac{\kappa^{i-1}d\kappa}{\left(\kappa+\alpha g-\beta\right)W}\right\} ,\label{eq:D(f,g)}
\end{eqnarray}
and $F_{i,k}\left(\kappa^{\left(1\right)},\ldots,\kappa^{\left(n\right)}\right)$
is a rational function of the $\kappa^{\left(j\right)},$ $j=1,\ldots n$.\\
Since (\ref{eq:ansatz_DlnZ_f}) and (\ref{eq:ansatz_DlnZ_g}) are
linear in $\partial_{f}u_{0}$ and $\partial_{g}u_{0}$, it is convenient
to restrict the range of possible $u_{0}$ to those, satisfying
\begin{alignat}{2}
0< & \lim_{\left(f,g\right)\rightarrow\left(\frac{1}{2},\frac{1}{2}\right)}\left[\left(\frac{1}{2}-g\right)\left|\partial_{g}u_{0}\right|^{2}\right] &  & =\gamma_{2}<\infty,\label{eq:11.23a-1-1}\\
0< & \lim_{\left(f,g\right)\rightarrow\left(\frac{1}{2},\frac{1}{2}\right)}\left[\left(\frac{1}{2}-f\right)\left|\partial_{f}u_{0}\right|^{2}\right] &  & =\gamma_{1}<\infty.\label{eq:11.23b-1-1}
\end{alignat}
Thus $u_{0}$ can be interpreted as a wave solution in its very own
right. Furthermore, the wave-solution corresponding to $u_{0}$ is
collinear, since $u_{0}$ is a real valued function, satisfying the
EPD equation.\\
However, we also have to examine the structure of (\ref{eq:A(f,g)})-(\ref{eq:D(f,g)})
for determining, whether $Z$ is satisfying (\ref{eq:boundary2})
and (\ref{eq:boundary2}). First of all, equation (\ref{eq:ansatz_DlnZ_f})
is considered together with condition (\ref{eq:11.23b-1-1}).\\
Because $u_{0}$ satisfies
\begin{eqnarray}
0< & \lim_{\left(f,g\right)\rightarrow\left(1/2,1/2\right)}\left[\left(\frac{1}{2}-f\right)\left|\partial_{f}u_{0}\right|^{2}\right] & =\gamma_{1}<\infty,\label{eq:u0_cond}
\end{eqnarray}
it follows that $\partial_{f}u_{0}$ is locally of the form
\begin{eqnarray*}
\partial_{f}u_{0}\left(f,\frac{1}{2}\right) & = & a+\frac{b}{\sqrt{1-2f}}+\mathcal{O}\left(X\right)\quad\textrm{close to }\left(f,g\right)=\left(\frac{1}{2},\frac{1}{2}\right),
\end{eqnarray*}
where $a,b\in\mathbb{\mathbb{C}}$ and $X$ is representative for
higher order terms in $f$ . Hence, $u_{0}$ assumes the local form
\begin{eqnarray*}
u_{0}\left(f,\frac{1}{2}\right) & = & af-\frac{b}{2}\sqrt{1-2f}+c+\mathcal{O}\left(X\right)\quad\textrm{close to }\left(f,g\right)=\left(\frac{1}{2},\frac{1}{2}\right),
\end{eqnarray*}
where $c\in\mathbb{\mathbb{C}}$. In addition, all the $u_{j}$, derived
from $u_{0}$ via relation (\ref{eq:rec_uj-1}), are at least of the
same order as $u_{0}$ close to $\left(f,g\right)=\left(\frac{1}{2},\frac{1}{2}\right)$,
since they are obtained by integrating (\ref{eq:rec_uj-1}) 
\begin{eqnarray}
u_{j}\left(f,g\right) & = & \frac{1}{2}\intop_{f_{0}}^{f}\left[\alpha u_{j-1}\left(\tilde{f},g\right)+2\left(\alpha\tilde{f}+\beta\right)\partial_{\tilde{f}}u_{j-1}\left(\tilde{f},g\right)\right]d\tilde{f}+v_{j}\left(g\right),\label{eq:uj_int}
\end{eqnarray}
with some arbitrary constant $f_{0}$ and a function $v_{j}\left(g\right)$
that needs to be determined by plugging $u_{j}\left(f,g\right)$ into
the EPD equation. As a result, the term $\frac{\alpha}{2}\sum_{l=1}^{i-1}\left(\alpha f+\beta\right)^{l-1}u_{i-l-1}$
in (\ref{eq:B(f,g)}) does not diverge for $\left(f,g\right)\rightarrow\left(\frac{1}{2},\frac{1}{2}\right)$
and the possibility 
\begin{eqnarray*}
\frac{\alpha}{2}\sum_{l=1}^{i-1}\left(\alpha f+\beta\right)^{l-1}u_{i-l-1} & \overset{\left(f,g\right)\rightarrow\left(\frac{1}{2},\frac{1}{2}\right)}{\longrightarrow} & 0
\end{eqnarray*}
can be excluded by choosing $\alpha$ and $\beta$ appropriately.\\
The next step is to exclude possible roots or poles of the term $\left(\kappa^{\left(k\right)}-\alpha f-\beta\right)\left(\kappa^{\left(k\right)}\right)^{n-1}$,
by noting that it is always possible to ensure $0<\left|\kappa^{\left(k\right)}-\alpha f-\beta\right|\left|\kappa^{\left(k\right)}\right|^{n-1}<\infty$
for $\left(f,g\right)\rightarrow\left(\frac{1}{2},\frac{1}{2}\right)$
by continuously changing the constants $\kappa_{k}$, $\alpha$ and
$\beta$. This follows since $\kappa^{\left(k\right)}$ is a nontrivial
hyperelliptic function of $u_{k}$ and thus of $\left(f,g\right)$,
which has a discrete set of poles and roots containing no limit points.
Therefore the location of poles and roots can be shifted by a continuous
variation of $\kappa_{k}$, such that there are no problems when $\left(f,g\right)\rightarrow\left(\frac{1}{2},\frac{1}{2}\right)$.
The possibility for $\sum_{m=1}^{n}\intop_{\kappa_{m}}^{\kappa^{\left(m\right)}}\frac{\kappa^{i-1}d\kappa}{\left(\kappa-\alpha f-\beta\right)W}$
to become singular is ruled out by the same argument.\\
Finally, $F_{i,k}\left(\kappa^{\left(1\right)},\ldots,\kappa^{\left(n\right)}\right)$
has no singular points, if all $\kappa^{\left(i\right)}$ are assumed
to be pairwise disjoint for each point $\left(f,g\right)$ in the
domain of consideration (consult the appendix for the detailed structure
of $F_{i,k}\left(\kappa^{\left(1\right)},\ldots,\kappa^{\left(n\right)}\right)$).\\
Accordingly, condition (\ref{eq:boundary2}) becomes
\begin{eqnarray}
\frac{1}{4} & \leq & \lim_{\left(f,g\right)\rightarrow\left(\frac{1}{2},\frac{1}{2}\right)}\left[\left(\frac{1}{2}-f\right)\frac{Z_{f}\bar{Z}_{f}}{\left(Z+\bar{Z}\right)^{2}}\right]\nonumber \\
 & = & \lim_{\left(f,g\right)\rightarrow\left(\frac{1}{2},\frac{1}{2}\right)}\left[\left(\frac{1}{2}-f\right)Z\bar{Z}\frac{\left(A\partial_{f}u_{0}+B\right)\left(\bar{A}\partial_{f}\bar{u}_{0}+\bar{B}\right)}{\left(Z+\bar{Z}\right)^{2}}\right]\nonumber \\
 & = & \frac{\left|A\left(\frac{1}{2},\frac{1}{2}\right)\right|^{2}\left|Z\left(\frac{1}{2},\frac{1}{2}\right)\right|^{2}\gamma_{1}}{4\left[\Re Z\left(\frac{1}{2},\frac{1}{2}\right)\right]^{2}}<\frac{1}{2},\label{eq:cond_hyperelliptic_f}
\end{eqnarray}
where (\ref{eq:u0_cond}) has been used. The requirement (\ref{eq:cond_hyperelliptic_f})
can be satisfied by replacing $u_{0}\rightarrow\delta u_{0}$ with
$\delta\in\mathbb{R}$, if $0<\left|A\left(\frac{1}{2},\frac{1}{2}\right)\right|<\infty$
and $0<\left|\Re Z\left(\frac{1}{2},\frac{1}{2}\right)\right|<\infty$,
since the EPD equation is invariant under a rescaling of $u_{0}$
with an arbitrary real parameter. Furthermore, the term $Z\left(\frac{1}{2},\frac{1}{2}\right)$
cannot cause any trouble, because neither $\sum_{m=1}^{n}\intop_{\kappa_{m}}^{\kappa^{\left(m\right)}}\frac{\kappa^{n}d\kappa}{W}$
nor $u_{n}$ are permitted to become singular for $\left(f,g\right)\rightarrow\left(\frac{1}{2},\frac{1}{2}\right)$
for an adequate choice of the constants $\kappa_{j}$. Similarly,
the case $A\left(\frac{1}{2},\frac{1}{2}\right)=0$ has already been
excluded. Therefore, the first condition (\ref{eq:boundary2}) can
always be satisfied.\\
The discussion of the second condition is in principle similar, when
assuming that the first condition has not already been fixed. However,
the situation turns out be highly nontrivial, since both conditions
need to be satisfied simultaneously. This can be seen as follows.\\
Plugging (\ref{eq:ansatz_DlnZ_g}) into the condition (\ref{eq:boundary1})
leads to the requirement
\begin{eqnarray}
\frac{1}{4} & \leq & \lim_{\left(f,g\right)\rightarrow\left(\frac{1}{2},\frac{1}{2}\right)}\left[\left(\frac{1}{2}-g\right)\frac{Z_{g}\bar{Z}_{g}}{\left(Z+\bar{Z}\right)^{2}}\right]\nonumber \\
 & = & \lim_{\left(f,g\right)\rightarrow\left(\frac{1}{2},\frac{1}{2}\right)}\left[\left(\frac{1}{2}-g\right)Z\bar{Z}\frac{\left(C\partial_{g}u_{0}+D\right)\left(\bar{C}\partial_{g}\bar{u}_{0}+\bar{D}\right)}{\left(Z+\bar{Z}\right)^{2}}\right]\nonumber \\
 & = & \frac{\left|C\left(\frac{1}{2},\frac{1}{2}\right)\right|^{2}\left|Z\left(\frac{1}{2},\frac{1}{2}\right)\right|^{2}\gamma_{2}}{4\left[\Re Z\left(\frac{1}{2},\frac{1}{2}\right)\right]^{2}}<\frac{1}{2},\label{eq:cond_hyperelliptic_g}
\end{eqnarray}
which appears to be problematic due to the obvious lack of symmetry
between $\partial_{f}\ln Z$ and $\partial_{g}\ln Z$, leading to
different values of $A\left(\frac{1}{2},\frac{1}{2}\right)$ and $C\left(\frac{1}{2},\frac{1}{2}\right)$.
Rescaling $u_{0}$ a second time to satisfy condition (\ref{eq:cond_hyperelliptic_g})
would also alter condition (\ref{eq:cond_hyperelliptic_f}). All in
all, it is not possible to give a general statement about the compliance
of $Z$ with the conditions (\ref{eq:cond_hyperelliptic_g}) and (\ref{eq:cond_hyperelliptic_f})
for arbitrary functions $u_{j}$, since the hyperelliptic functions
do not admit an explicit evaluation of $A\left(f,g\right)$, $B\left(f,g\right)$,
$C\left(f,g\right)$, and $D\left(f,g\right)$ at $f=\frac{1}{2}$
and $g=\frac{1}{2}$. In terms of degrees of freedom, there are $2n+3$
real parameters from the $n$ complex constants $\kappa_{i}$ ($i=1,2,\ldots,n$),
$\alpha$, and $\beta$ plus a possible rescaling of $u_{0}$, which
can be adjusted in order to satisfy the two conditions (\ref{eq:cond_hyperelliptic_g})
and (\ref{eq:cond_hyperelliptic_f}). However, each specific case
has to be considered separately, since the explicit dependencies of
$A\left(f,g\right)$, $B\left(f,g\right)$, $C\left(f,g\right)$,
and $D\left(f,g\right)$ on those parameters are unknown. An expansion
of the regarding hyperelliptic functions to a power series close to
$f=\frac{1}{2}$ and $g=\frac{1}{2}$ might give a deeper insight
on their local dependency on the $\kappa_{i}$ ($i=1,2,\ldots,n$)
, $\alpha$, $\beta$ and $u_{0}$, such that a suitable choice of
these parameters might become easier and a general statement can be
formulated.

\noindent Note, the relaxed wave conditions (cf. \cite{griffiths1991colliding})
\begin{alignat}{2}
0\leq & \lim_{\left(f,g\right)\rightarrow\left(\frac{1}{2},\frac{1}{2}\right)}\left[\left(\frac{1}{2}-g\right)\frac{Z_{g}\bar{Z}_{g}}{\left(Z+\bar{Z}\right)^{2}}\right] &  & =\frac{k_{2}}{2}<\frac{1}{2},\label{eq:11.23a-2-1}\\
0\leq & \lim_{\left(f,g\right)\rightarrow\left(\frac{1}{2},\frac{1}{2}\right)}\left[\left(\frac{1}{2}-f\right)\frac{Z_{f}\bar{Z}_{f}}{\left(Z+\bar{Z}\right)^{2}}\right] &  & =\frac{k_{1}}{2}<\frac{1}{2},\label{eq:11.23b-2-1}
\end{alignat}
can always be met by rescaling $u_{0}$ with some constant $0<\delta<1$.
This possibility will be ignored here, because of the unclear physical
implications emerging from impulsive matter tensor components at the
junctions $f=\frac{1}{2}$ and $g=\frac{1}{2}$, which can occur in
this case.

\noindent One might ask, what possibilities remain for $Z$ to satisfy
the junction conditions (\ref{eq:boundary2})-(\ref{eq:boundary1}),
if $u_{0}$ does not fulfill (\ref{eq:u0_cond}). Possible singularities
of the right order may occur in the functions $A\left(f,g\right)$
and $B\left(f,g\right)$ (cf. (\ref{eq:A(f,g)})-(\ref{eq:B(f,g)})).
However, the junction conditions cannot be met, if both functions
only exhibit singularities of integer order, since the term $\frac{Z_{f}\bar{Z}_{f}}{\left(Z+\bar{Z}\right)^{2}}$
would then have a singularity of higher order than 1 and accordingly
$\lim_{\left(f,g\right)\rightarrow\left(\frac{1}{2},\frac{1}{2}\right)}\left[\left(\frac{1}{2}-f\right)\frac{Z_{f}\bar{Z}_{f}}{\left(Z+\bar{Z}\right)^{2}}\right]$
would not be finite. However, no general statement can be made for
arbitrary functions $u_{j}$.

\noindent All in all, the procedure presented here corresponds to
a particular simple way for generating arbitrary polarized solutions
from collinear ones, which cannot be reduced to a simple coordinate
transformation. Moreover, the solutions generated by this method are
highly non-trivial due to the occurrence of hyperelliptic integrals.
Meinel and Neugebauer have mentioned (\cite{meinel1996solutions}),
that it is generally possible to solve the inversion problem (\ref{eq:inversion_prob})
by means of hyperelliptic functions. Still, the analysis and interpretation
of solution in terms of these special functions remain a hard challenge
and their physical relevance might be doubtful.

\subsection*{An example: The hyperelliptic class for $n=1$}

\noindent For $n=1$, the regarding hyperelliptic integrals reduce
to elliptic integrals, that can be expressed in terms of elliptic
functions. The details of the calculation can be found in the appendix,
the result is an inherently non-linear solution $Z$ of the hyperbolic
Ernst equation (\ref{eq:f_g_hyp_ernst}):
\begin{eqnarray}
Z\left(f,g\right) & = & \exp\left\{ \pm\frac{2i\left(\bar{\kappa}_{1}-\kappa_{1}\right)\Pi\left[k\left(f,g\right),\textrm{am}\left(\tilde{u}_{0}\left(f,g\right),m\left(f,g\right)\right),m\left(f,g\right)\right]}{\sqrt{\left(\alpha f+\beta-\kappa_{1}\right)\left(\alpha g-\beta+\bar{\kappa}_{1}\right)}}\right.\nonumber \\
 &  & \left.\pm\bar{\kappa}_{1}u_{0}\left(f,g\right)-u_{1}\left(f,g\right)\right\} ,\label{eq:Z_from_u0-1}
\end{eqnarray}
where $\Pi\left(\cdot,\cdot,\cdot\right)$ denotes the incomplete
elliptic integral of the third kind, $\textrm{am\ensuremath{\left(\cdot,\cdot\right)}}$
is the Jacobian amplitude function (cf. \cite{abr70}) and $u_{1}\left(f,g\right)$
follows from (\ref{eq:uj_int}):
\begin{eqnarray*}
u_{1}\left(f,g\right) & = & \frac{1}{2}\intop_{f_{0}}^{f}\left[\alpha u_{0}\left(\tilde{f},g\right)+2\left(\alpha\tilde{f}+\beta\right)\partial_{\tilde{f}}u_{0}\left(\tilde{f},g\right)\right]d\tilde{f}+v_{1}\left(g\right),
\end{eqnarray*}
and the following abbreviations have been used 
\begin{eqnarray}
\tilde{u}_{0} & = & \mp\sqrt{\left(\kappa_{1}-\alpha f-\beta\right)\left(\bar{\kappa}_{1}+\alpha g-\beta\right)}\frac{u_{0}}{2},\label{eq:u0s}\\
m\left(f,g\right) & = & \frac{\left(\bar{\kappa}_{1}-\alpha f-\beta\right)\left(\kappa_{1}+\alpha g-\beta\right)}{\left(\kappa_{1}-\alpha f-\beta\right)\left(\bar{\kappa}_{1}+\alpha g-\beta\right)},\label{eq:m(f,g)}\\
k\left(f,g\right) & = & \frac{\left(\kappa_{1}+\alpha g-\beta\right)}{\left(\bar{\kappa}_{1}+\alpha g-\beta\right)}.\label{eq:k(f,g)}
\end{eqnarray}

\subsubsection*{An elliptic solution from a Khan-Penrose seed}

\begin{figure}[H]
\noindent \begin{centering}
\includegraphics[scale=0.8]{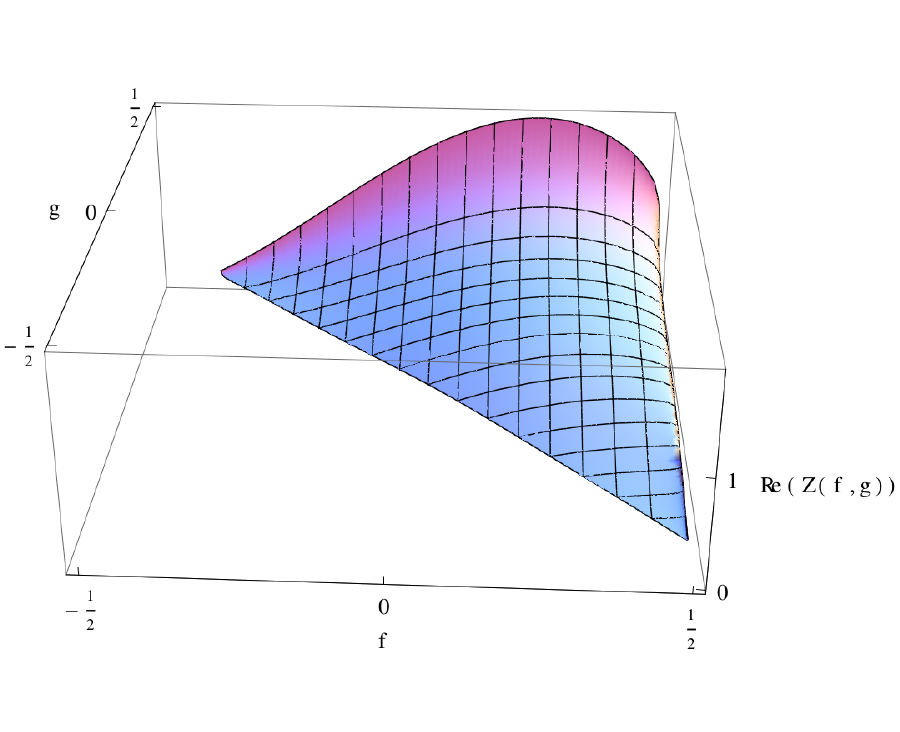}
\par\end{centering}

\caption{\label{fig:Real-part-of}Real part of $Z\left(f,g\right)$ given by
(\ref{eq:Z_from_u0-1-1}) in the domain $\Omega_{f,g}$.}

\end{figure}

\begin{figure}[H]
\noindent \begin{centering}
\includegraphics[scale=0.8]{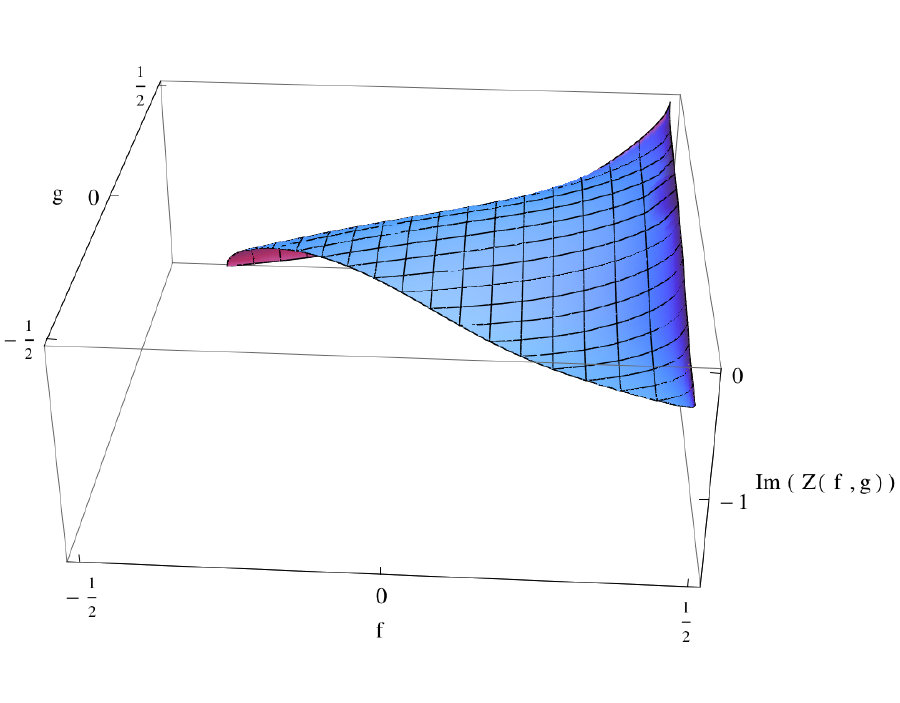}
\par\end{centering}

\caption{\label{fig:Imaginary-part-of}Imaginary part of $Z\left(f,g\right)$
given by (\ref{eq:Z_from_u0-1-1}) in the domain $\Omega_{f,g}$.}
\end{figure}

\noindent For demonstrating the practical relevance of the methods
discussed so far, a special solution for the $n=1$ case is presented,
where the well-known Khan Penrose solution (cf. \cite{khan1971scattering})
is taken as a seed solution $u_{0}$:

\begin{eqnarray}
u_{0} & = & 2\gamma\textrm{arctanh}\left(\sqrt{\frac{1}{2}-f}\sqrt{\frac{1}{2}+g}-\sqrt{\frac{1}{2}+f}\sqrt{\frac{1}{2}-g}\right)\label{eq:u0_Khan_Penrose-1}
\end{eqnarray}
where $\gamma$ is used for fitting the generated solution to the
junction conditions.\\
After trying some different values for $\alpha$, $\beta$ and $\kappa_{1}$,
the following choice turns out to be compatible with the wave conditions
\begin{alignat}{6}
\alpha & = & 1, & \quad & \beta & = & 0, & \quad & \kappa_{1} & = & \frac{i}{4},\label{eq:choice_const}
\end{alignat}
yielding 
\begin{eqnarray}
u_{1}\left(f,g\right) & = & \frac{\gamma}{2}\left\{ \sqrt{1+2f}\sqrt{1-2g}-\sqrt{1+2g}\sqrt{1-2f}\right.\nonumber \\
 &  & \left.-2\left(f-g\right)\textrm{arctanh}\left[\frac{1}{2}\left(\sqrt{1+2f}\sqrt{1-2g}+\sqrt{1+2g}\sqrt{1-2f}\right)\right]\right\} \label{eq:u1_Khan-1}
\end{eqnarray}
which is a valid solution of the EPD equation that satisfies (\ref{eq:rec_uj-1})-(\ref{eq:rec_uj2-1}).\\
Hence, the following solution has been generated
\begin{eqnarray}
Z\left(f,g\right) & = & \exp\left\{ \frac{i}{\sqrt{\left(f-\frac{i}{4}\right)\left(g-\frac{i}{4}\right)}}\Pi\left[k\left(f,g\right),\textrm{am}\left(-\tilde{u}_{0}\left(f,g\right),m\left(f,g\right)\right),m\left(f,g\right)\right]\right.\nonumber \\
 &  & \left.+\frac{i}{4}u_{0}\left(f,g\right)-u_{1}\left(f,g\right)\right\} ,\label{eq:Z_from_u0-1-1}
\end{eqnarray}
where 
\begin{eqnarray}
\tilde{u}_{0} & = & \sqrt{\left(\frac{i}{4}-f\right)\left(g-\frac{i}{4}\right)}\frac{u_{0}}{2},\label{eq:u0s_specific}
\end{eqnarray}
and all signs have been chosen appropriately, such that the underlying
square-roots are taken consistently.\\
Furthermore, choosing $\gamma=2$ gives the following limit values
\begin{eqnarray*}
\lim_{\left(f,g\right)\rightarrow\left(\frac{1}{2},\frac{1}{2}\right)}\left[\sqrt{\frac{1}{2}-g}\frac{Z_{g}}{Z}\right] & = & 1+\frac{i}{2},\\
\lim_{\left(f,g\right)\rightarrow\left(\frac{1}{2},\frac{1}{2}\right)}\left[\sqrt{\frac{1}{2}-f}\frac{Z_{f}}{Z}\right] & = & -1+\frac{i}{2},\\
\lim_{\left(f,g\right)\rightarrow\left(\frac{1}{2},\frac{1}{2}\right)}Z & = & 1,
\end{eqnarray*}
leading to
\begin{eqnarray}
\lim_{\left(f,g\right)\rightarrow\left(\frac{1}{2},\frac{1}{2}\right)}\left[\left(\frac{1}{2}-g\right)\frac{Z_{g}\bar{Z}_{g}}{\left(Z+\bar{Z}\right)^{2}}\right] & = & \frac{5}{16},\label{eq:limit1}\\
\lim_{\left(f,g\right)\rightarrow\left(\frac{1}{2},\frac{1}{2}\right)}\left[\left(\frac{1}{2}-f\right)\frac{Z_{f}\bar{Z}_{f}}{\left(Z+\bar{Z}\right)^{2}}\right] & = & \frac{5}{16}.\label{eq:limit2}
\end{eqnarray}
Thus, (\ref{eq:Z_from_u0-1-1}) satisfies the junction conditions
(\ref{eq:boundary2})-(\ref{eq:boundary1}). As a result, a potential
$Z$ describing a new colliding plane wave space-time has been created.
Figures \ref{fig:Real-part-of}-\ref{fig:Imaginary-part-of} show
the real and imaginary parts of $Z$ in the domain $\Omega_{f,g}=\left\{ \left(f,g\right)\in\mathbb{R}^{2},\; f<\frac{1}{2},\; g<\frac{1}{2},\; f+g>0\right\} $.


\section{Conclusions and outlook}

As a main result, a new solution class for the hyperbolic Ernst equation
has been obtained. Furthermore, this class offers a way for creating
non-parallel polarized wave solutions from collinear wave solutions.
Consequently, it has been shown that solutions of the EPD equation
that satisfy the wave conditions can be used as seed solutions for
generating general polarized waves related to a hyperelliptic class
of solutions of the Ernst equation. In this context, a specific example
of an elliptic solution generated from the Khan-Penrose solution has
been provided.\\
However, it has not been possible to formulate a general argument,
delivering an exact mathematical statement about the compliance of
solutions obtained from arbitrary seeds with the wave conditions emerging
in relation to plane-wave collisions. A series expansion of the regarding
hyperelliptic functions with respect to the external parameters could
be one possibility for obtaining the desired statement. Obviously,
this problem leaves space for further research.\\
Moreover, analytical features of the new hyperelliptic solution class
can be examined by considering more specific examples, of which a
possible wave analogue for the stationary rotating disk of dust (see
\cite{meinel1996solutions}) might be of particular importance.\\
All together, the complexity of the obtained solution class opens
a wide field for further research also in pure mathematics and of
course for colliding plane-wave space-times.

\appendix

\section*{Appendix}

\subsection*{Computing the first derivatives}

\noindent In this section, the reduction of $\frac{\partial_{f}\kappa^{\left(m\right)}}{W^{\left(m\right)}}$
into terms of the $u_{j}$ and corresponding derivatives is performed.
For this purpose, (\ref{eq:Dz_uj-1}) is written in the form
\begin{eqnarray}
\sum_{m=1}^{n}\frac{\partial_{f}\kappa^{\left(m\right)}}{W^{\left(m\right)}}\left(\kappa^{\left(m\right)}\right)^{j} & = & \partial_{f}u_{j}-\frac{\alpha}{2}\sum_{m=1}^{n}\intop_{\kappa_{m}}^{\kappa^{\left(m\right)}}\frac{\kappa^{j}d\kappa}{\left(\kappa-\alpha f-\beta\right)W},\quad j=0,1,2,\ldots,n-1,\nonumber \\
\mathbf{v}^{T}\mathbf{A} & = & \mathbf{w},\label{eq:matrix_form}
\end{eqnarray}
where the vectors $\mathbf{v}$, $\mathbf{w}\in\mathbb{C}^{n}$ and
the matrix $\mathbf{A}\in\mathbb{C}^{n\times n}$ have been introduced
according to
\begin{alignat}{3}
\left(\mathbf{v}\right)_{i} & = &  & \frac{\partial_{f}\kappa^{\left(i\right)}}{W^{\left(i\right)}} & \quad & i=1,2,\ldots,n,\label{eq:vector_v}\\
\left(\mathbf{w}\right)_{j} & = &  & \partial_{f}u_{j-1}-\frac{\alpha}{2}\sum_{m=1}^{n}\intop_{\kappa_{m}}^{\kappa^{\left(m\right)}}\frac{\kappa^{j-1}d\kappa}{\left(\kappa-\alpha f-\beta\right)W} & \qquad & j=1,2,\ldots,n,\label{eq:vector_w}\\
\left(\mathbf{A}\right)_{ij} & = &  & \left(\kappa^{\left(i\right)}\right)^{j-1}. & \quad\label{eq:vector_A}
\end{alignat}
Hence, $\mathbf{A}$ is a Vandermonde matrix
\begin{eqnarray}
\mathbf{A} & = & \left(\begin{array}{cccc}
1 & \kappa^{\left(1\right)} & \cdots & \left(\kappa^{\left(1\right)}\right)^{n-1}\\
1 & \kappa^{\left(2\right)} & \ldots & \left(\kappa^{\left(2\right)}\right)^{n-1}\\
\vdots & \vdots & \ddots & \vdots\\
1 & \kappa^{\left(n\right)} & \ldots & \left(\kappa^{\left(n\right)}\right)^{n-1}
\end{array}\right),\label{eq:vandermonde_A}
\end{eqnarray}
that can be inverted, if
\begin{alignat}{3}
\det\mathbf{A} & = & \prod_{1\leq i<j\leq n}\left(\kappa^{\left(j\right)}-\kappa^{\left(i\right)}\right) & \neq & 0.\label{eq:detA}
\end{alignat}
Therefore, the $\kappa^{\left(i\right)}$ are assumed to be pairwise
disjoint for each point $\left(f,g\right)$ in the domain of consideration.\\
The inverse of $\mathbf{A}$ reads
\begin{eqnarray}
\left(\mathbf{A}^{-1}\right)_{ij} & = & \left(-1\right)^{i}\left\{ \frac{{\displaystyle \sum_{\begin{array}{c}
{\scriptstyle 1\leq m_{1}<\ldots<m_{n-i+1}\leq n}\\
{\scriptstyle m_{1},\ldots,m_{n-i+1}\neq j}
\end{array}}\kappa^{\left(m_{1}\right)}\cdots\kappa^{\left(m_{n-i+1}\right)}}}{{\displaystyle \prod_{\begin{array}{c}
{\scriptstyle 1\leq m\leq n}\\
{\scriptstyle m\neq j}
\end{array}}}\left(\kappa^{\left(m\right)}-\kappa^{\left(j\right)}\right)}\right\} ,\label{eq:inverse_A}
\end{eqnarray}
and accordingly $\frac{\partial_{f}\kappa^{\left(j\right)}}{W^{\left(j\right)}}$
can be written as
\begin{eqnarray}
\frac{\partial_{f}\kappa^{\left(j\right)}}{W^{\left(j\right)}} & = & \sum_{i=1}^{n}\left(\mathbf{w}\right)_{i}\left(\mathbf{A}^{-1}\right)_{ij}\nonumber \\
 & = & \sum_{i=1}^{n}F_{ij}\left(\kappa^{\left(1\right)},\ldots,\kappa^{\left(n\right)}\right)\left\{ \partial_{f}u_{i-1}-\frac{\alpha}{2}\sum_{m=1}^{n}\intop_{\kappa_{m}}^{\kappa^{\left(m\right)}}\frac{\kappa^{i-1}d\kappa}{\left(\kappa-\alpha f-\beta\right)W}\right\} ,\label{eq:Dz_kappa}
\end{eqnarray}
where $F_{ij}\left(\kappa^{\left(1\right)},\ldots,\kappa^{\left(n\right)}\right)=\left(\mathbf{A}^{-1}\right)_{ij}$
is a rational function of $\kappa^{\left(1\right)},\ldots,\kappa^{\left(n\right)}$.
Similarly, the derivative with respect to $g$ leads to
\begin{eqnarray}
\frac{\partial_{g}\kappa^{\left(j\right)}}{W^{\left(j\right)}} & = & \sum_{i=1}^{n}F_{ij}\left(\kappa^{\left(1\right)},\ldots,\kappa^{\left(n\right)}\right)\left\{ \partial_{g}u_{i-1}+\frac{\alpha}{2}\sum_{m=1}^{n}\intop_{\kappa_{m}}^{\kappa^{\left(m\right)}}\frac{\kappa^{i-1}d\kappa}{\left(\kappa+\alpha g-\beta\right)W}\right\} .\label{eq:Dzs_kappa}
\end{eqnarray}
As a result, the derivatives $\partial_{f}\ln Z$ and $\partial_{g}\ln Z$
can be written in terms of the $u_{j}$, when plugging (\ref{eq:Dz_kappa})
and (\ref{eq:Dzs_kappa}) into (\ref{eq:Dz_lnZ-1}) and (\ref{eq:Dzs_lnZ-1})
\begin{eqnarray}
\partial_{f}\ln Z & = & \sum_{k,i=1}^{n}\left(\kappa^{\left(k\right)}-\alpha f-\beta\right)\left(\kappa^{\left(k\right)}\right)^{n-1}F_{ik}\left(\kappa^{\left(1\right)},\ldots,\kappa^{\left(n\right)}\right)\nonumber \\
 &  & \times\left\{ \partial_{f}u_{i-1}-\frac{\alpha}{2}\sum_{m=1}^{n}\intop_{\kappa_{m}}^{\kappa^{\left(m\right)}}\frac{\kappa^{i-1}d\kappa}{\left(\kappa-\alpha f-\beta\right)W}\right\} ,\label{eq:Dz_lnZ_u}\\
\partial_{g}\ln Z & = & \sum_{k,i=1}^{n}\left(\kappa^{\left(k\right)}+\alpha g-\beta\right)\left(\kappa^{\left(k\right)}\right)^{n-1}F_{ik}\left(\kappa^{\left(1\right)},\ldots,\kappa^{\left(n\right)}\right)\nonumber \\
 &  & \times\left\{ \partial_{g}u_{i-1}+\frac{\alpha}{2}\sum_{m=1}^{n}\intop_{\kappa_{m}}^{\kappa^{\left(m\right)}}\frac{\kappa^{i-1}d\kappa}{\left(\kappa+\alpha g-\beta\right)W}\right\} .\label{eq:Dzs_lnZ_u}
\end{eqnarray}
The next step is to resolve the recursive relation (\ref{eq:rec_uj})
in order to shrink down the term $\partial_{f}u_{i-1}$ to its basic
ingredients
\begin{eqnarray*}
\partial_{f}u_{j} & = & \frac{\alpha}{2}u_{j-1}+\left(\alpha f+\beta\right)\underbrace{\partial_{f}u_{j-1}}_{=\frac{\alpha}{2}u_{j-2}+\left(\alpha f+\beta\right)\partial_{f}u_{j-2}}\\
 & = & \frac{\alpha}{2}\left(u_{j-1}+\left(\alpha f+\beta\right)u_{j-2}\right)+\left(\alpha f+\beta\right)^{2}\underbrace{\partial_{z}u_{j-2}}_{=\frac{\alpha}{2}u_{j-3}+\left(\alpha f+\beta\right)\partial_{f}u_{j-3}}\\
 & = & \frac{\alpha}{2}\left(u_{j-1}+\left(\alpha f+\beta\right)u_{j-2}+\left(\alpha f+\beta\right)^{2}u_{j-3}\right)+\left(\alpha f+\beta\right)^{3}\partial_{f}u_{j-3}\\
 & \vdots\\
 & = & \frac{\alpha}{2}\sum_{k=1}^{j}\left(\alpha f+\beta\right)^{k-1}u_{j-k}+\left(\alpha f+\beta\right)^{j}\partial_{f}u_{0}.
\end{eqnarray*}
A similar relation holds for $\partial_{g}u_{j}$. Consequently, (\ref{eq:Dz_lnZ_u})
and (\ref{eq:Dzs_lnZ_u}) become
\begin{eqnarray}
\partial_{f}\ln Z & = & \sum_{k,i=1}^{n}\left(\kappa^{\left(k\right)}-\alpha f-\beta\right)\left(\kappa^{\left(k\right)}\right)^{n-1}F_{ik}\left(\kappa^{\left(1\right)},\ldots,\kappa^{\left(n\right)}\right)\nonumber \\
 &  & \times\left\{ \frac{\alpha}{2}\sum_{l=1}^{i-1}\left(\alpha f+\beta\right)^{l-1}u_{i-l-1}+\left(\alpha f+\beta\right)^{i-1}\partial_{f}u_{0}\right.\nonumber \\
 &  & \left.-\frac{\alpha}{2}\sum_{m=1}^{n}\intop_{\kappa_{m}}^{\kappa^{\left(m\right)}}\frac{\kappa^{i-1}d\kappa}{\left(\kappa-\alpha f-\beta\right)W}\right\} ,\label{eq:Dz_lnZ_u-1}\\
\partial_{\bar{z}}\ln Z & = & \sum_{k,i=1}^{n}\left(\kappa^{\left(k\right)}+\alpha g-\beta\right)\left(\kappa^{\left(k\right)}\right)^{n-1}F_{ik}\left(\kappa^{\left(1\right)},\ldots,\kappa^{\left(n\right)}\right)\nonumber \\
 &  & \times\left\{ -\frac{\alpha}{2}\sum_{l=1}^{i-1}\left(\beta-\alpha g\right)^{l-1}u_{i-l-1}+\left(\beta-\alpha g\right)^{i-1}\partial_{g}u_{0}\right.\nonumber \\
 &  & \left.+\frac{\alpha}{2}\sum_{m=1}^{n}\intop_{\kappa_{m}}^{\kappa^{\left(m\right)}}\frac{\kappa^{i-1}d\kappa}{\left(\kappa+\alpha g-\beta\right)W}\right\} ,\label{eq:Dzs_lnZ_u-1}
\end{eqnarray}
which shows that $\partial_{f}\ln Z$ and $\partial_{g}\ln Z$ are
linear functions of $\partial_{f}u_{0}$ and $\partial_{g}u_{0}$,
respectively.

\subsection*{The elliptic case}

\noindent The hyperelliptic solution (\ref{eq:meinel_sol_hyper})
reduces to the following form for $n=1$
\begin{eqnarray}
Z & = & \exp\left(\intop_{\kappa_{1}}^{\kappa^{\left(1\right)}}\frac{\kappa d\kappa}{W\left(\kappa\right)}-u_{1}\right),\label{eq:meinel_hyperelliptic1-1-1-1-1}
\end{eqnarray}
where 
\begin{eqnarray}
W\left(\kappa;f,g\right){}^{2} & = & \left(\kappa-\alpha f-\beta\right)\left(\kappa+\alpha g-\beta\right)\left(\kappa-\kappa_{1}\right)\left(\kappa-\bar{\kappa}_{1}\right),\label{eq:riemann_W-1-1-1-1}
\end{eqnarray}
and $\kappa^{\left(1\right)}$ is the solution of the following inversion
problem
\begin{eqnarray}
\intop_{\kappa_{1}}^{\kappa^{\left(1\right)}}\frac{d\kappa}{W\left(\kappa\right)} & = & u_{0}.\label{eq:inversion_prob-1-1-1-1}
\end{eqnarray}
Writing $W^{2}$ a bit more general 
\begin{eqnarray}
W^{2}\left(\kappa\right) & = & \left(\kappa-e_{1}\right)\left(\kappa-e_{2}\right)\left(\kappa-e_{3}\right)\left(\kappa-e_{4}\right),\label{eq:W_e1_e2_e3_e4-1}
\end{eqnarray}
yields the following primitive for the integral (\ref{eq:inversion_prob-1-1-1-1})
\begin{eqnarray}
\intop\frac{d\kappa}{W\left(\kappa\right)} & = & \pm\frac{2}{\sqrt{\left(e_{1}-e_{4}\right)\left(e_{3}-e_{2}\right)}}\textrm{F}\left[\arcsin\sqrt{\Lambda\left(\kappa\right)},m\right],\label{eq:int_1/W-2}
\end{eqnarray}
where $F\left(\cdot,\cdot\right)$ is the incomplete elliptic integral
of the first kind (see \cite{abr70}) and 
\begin{eqnarray}
\Lambda\left(\kappa\right) & = & \frac{\left(e_{2}-e_{3}\right)\left(e_{4}-\kappa\right)}{\left(e_{4}-e_{3}\right)\left(e_{2}-\kappa\right)},\label{eq:Lambda-1}\\
m & = & \frac{\left(e_{2}-e_{1}\right)\left(e_{4}-e_{3}\right)}{\left(e_{4}-e_{1}\right)\left(e_{2}-e_{3}\right)}.\label{eq:m-1}
\end{eqnarray}
The sign ambiguity in (\ref{eq:int_1/W-2}) is a reminder that some
square-roots have been taken, which should have been treated with
more care (especially when dealing with specific values of $\kappa_{1}$).
The same holds for the following sign ambiguities.\\
Choosing $e_{4}=\kappa_{1}$, $e_{2}=\bar{\kappa}_{1}$, $e_{1}=\alpha f+\beta$
and $e_{3}=\beta-\alpha g$ leads to 
\begin{alignat}{3}
u_{0} & = & \intop_{\kappa_{1}}^{\kappa}\frac{d\kappa^{\prime}}{W\left(\kappa^{\prime}\right)} & = & \pm\frac{2}{\sqrt{\left(e_{1}-e_{4}\right)\left(e_{3}-e_{2}\right)}}\textrm{F}\left[\arcsin\sqrt{\Lambda\left(\kappa\right)},m\right],\label{eq:u0_F-1-1}
\end{alignat}
because $\textrm{F}\left[0,m\right]=0$ and therefore
\begin{eqnarray}
\arcsin\sqrt{\Lambda\left(\kappa\right)} & = & \textrm{am}\left(\tilde{u}_{0},m\right),\label{eq:arcsin(sqrt(Lambda))-1}\\
\sqrt{\Lambda\left(\kappa\right)} & = & \textrm{sn}\left(\tilde{u}_{0},m\right),\label{eq:sqrt(Lambda)-1}
\end{eqnarray}
where $\textrm{am\ensuremath{\left(\cdot,\cdot\right)}}$ denotes
the amplitude function and $\textrm{sn\ensuremath{\left(\cdot,\cdot\right)}}$
is the Jacobian elliptic sine amplitude (see \cite{abr70}) and the
following abbreviation has been introduced 
\begin{eqnarray}
\tilde{u}_{0} & = & \pm\sqrt{\left(e_{1}-e_{4}\right)\left(e_{3}-e_{2}\right)}\frac{u_{0}}{2}.\label{eq:u0_schlange-1-1}
\end{eqnarray}
The primitive for the integral occurring in (\ref{eq:meinel_hyperelliptic1-1-1-1-1})
is
\begin{eqnarray}
\intop\frac{\kappa d\kappa}{W\left(\kappa\right)} & = & \pm\frac{2}{\sqrt{\left(e_{1}-e_{4}\right)\left(e_{3}-e_{2}\right)}}\left\{ e_{2}\textrm{F}\left[\arcsin\sqrt{\Lambda\left(\kappa\right)},m\right]\right.\nonumber \\
 &  & \left.+\left(e_{2}-e_{4}\right)\Pi\left[k,\arcsin\sqrt{\Lambda\left(\kappa\right)},m\right]\right\} \label{eq:int_k/W-3-1}
\end{eqnarray}
where $\Pi\left(\cdot,\cdot,\cdot\right)$ denotes the incomplete
elliptic integral of the third kind (cf. \cite{abr70}) and
\begin{eqnarray}
k & = & \frac{\left(e_{4}-e_{3}\right)}{\left(e_{2}-e_{3}\right)}.\label{eq:k_e2_e3_e4-1}
\end{eqnarray}
Hence, the regarding definite integral reads
\begin{eqnarray}
\intop_{\kappa_{1}}^{\kappa}\frac{\kappa d\kappa}{W} & = & \left\{ e_{2}u_{0}\pm\frac{2\left(e_{2}-e_{4}\right)}{\sqrt{\left(e_{1}-e_{4}\right)\left(e_{3}-e_{2}\right)}}\Pi\left[k,\textrm{am}\left(\tilde{u}_{0},m\right),m\right]\right\} ,\label{eq:int_k/W-4-1}
\end{eqnarray}
since $\Pi\left[k,0,m\right]=0$.\\
Putting everything together, the resulting Ernst potential $Z$ reads:
\begin{eqnarray}
Z\left(f,g\right) & = & \exp\left\{ \pm\frac{2i\left(\bar{\kappa}_{1}-\kappa_{1}\right)}{\sqrt{\left(\alpha f+\beta-\kappa_{1}\right)\left(\alpha g-\beta+\bar{\kappa}_{1}\right)}}\Pi\left[k\left(f,g\right),\textrm{am}\left(\tilde{u}_{0}\left(f,g\right),m\left(f,g\right)\right),m\left(f,g\right)\right]\right.\nonumber \\
 &  & \left.\pm\bar{\kappa}_{1}u_{0}\left(f,g\right)-u_{1}\left(f,g\right)\right\} .\label{eq:Z_from_u0-1-2}
\end{eqnarray}

\bibliographystyle{plain}
\bibliography{paper_hyperelliptic}

\end{document}